\documentclass[twocolumn,showpacs,prb,amsmath,amssymb,floatfix,eqsecnum,subeqn]{revtex4}
\usepackage{amsmath,amssymb}
\usepackage{bm}
\usepackage{epsfig}
\usepackage{psfrag}

\newcommand{\bd}{\bm}
\newcommand{\jonecolor}{green }
\newcommand{\jtwocolor}{blue }

\begin{document}

\title{Magnetic properties of a metal-organic  antiferromagnet 
  on  a distorted honeycomb lattice}

\author{Ivan Spremo, Florian Sch\"{u}tz, and Peter Kopietz}  
\affiliation{Institut f\"{u}r Theoretische Physik, Universit\"{a}t
  Frankfurt, Max-von-Laue-Strasse 1, 60438 Frankfurt, Germany}

\author{Volodymyr Pashchenko, Bernd Wolf, and Michael Lang}
\affiliation{Physikalisches Institut, Universit\"{a}t
  Frankfurt, Max-von-Laue-Strasse 1, 60438 Frankfurt, Germany}

\author{Jan W.~Bats}
\affiliation{Institut f\"ur Organische Chemie und Chemische Biologie, 
  Universit\"{a}t Frankfurt, Marie-Curie-Strasse 11, 60439 Frankfurt, Germany}

\author{Chunhua Hu and Martin U.~Schmidt}
\affiliation{Institut f\"ur Anorganische und Analytische Chemie, 
  Universit\"{a}t Frankfurt, Marie-Curie-Strasse 11, 60439 Frankfurt, Germany}

\date{\today}

\begin{abstract}
  For temperatures $T$ well above the ordering temperature $T_{\ast}
  = 3.0 \pm 0.2 \mathrm{K}$ the magnetic properties of the metal-organic material 
  $\mathrm{Mn}[\mathrm{C}_{10}\mathrm{H}_{6}(\mathrm{OH})(\mathrm{COO})]_{2}
  \!\times\! 2\mathrm{H}_{2}\mathrm{O}$
  built from
  Mn$^{2+}$ ions and $3$-hydroxy-2-naphthoic anions can be described by a
  $S=5/2$ quantum antiferromagnet on a distorted honeycomb lattice
  with two different nearest neighbor exchange couplings $J_2 \approx
  2 J_1 \approx 1.8\, \mathrm{K}$.  Measurements of the magnetization
  $M(H,T)$ as a function of a uniform external field $H$ and of the
  uniform zero field susceptibility $\chi(T)$ are explained within the
  framework of a modified spin-wave approach which takes into account
  the absence of a spontaneous staggered magnetization at finite
  temperatures.
\end{abstract}




\maketitle

\section{Introduction}

In recent years the role of fluctuations, spatial anisotropy and
frustration in low dimensional quantum magnets has been intensely
studied, both experimentally and theoretically.\cite{Schollwoeck04}
For a comparison of experiments with theory it is crucial to have well
defined crystalline materials where one or several parameters can be
varied externally in order to obtain quantitative  predictions for
physical observables.  Moreover, in order to observe interesting
magnetic many-body effects it is essential to have materials where the
magnetic moments are coupled via sufficiently strong exchange
interactions.  These conditions are met by transition metal oxides
such as cuprates, vanadates, copper-germanates, or manganites, which
have been the subject of many works.  However, in these materials it
is rather difficult to control externally microscopic parameters such
as the precise values of the exchange interactions or the lattice geometry by
changing the chemical composition in a well defined manner.  This
problem tends to be less severe in magnets based on metal-organic
materials, which offer more possibilities of modifying some
constituents chemically and thereby tuning the properties by a crystal 
engineering strategy.  
The challenge is then to find metal-organic
magnets where the magnetic moments are coupled sufficiently strongly
to exhibit interesting collective effects.

\begin{figure}[tb]
  \begin{center}
    \epsfig{file=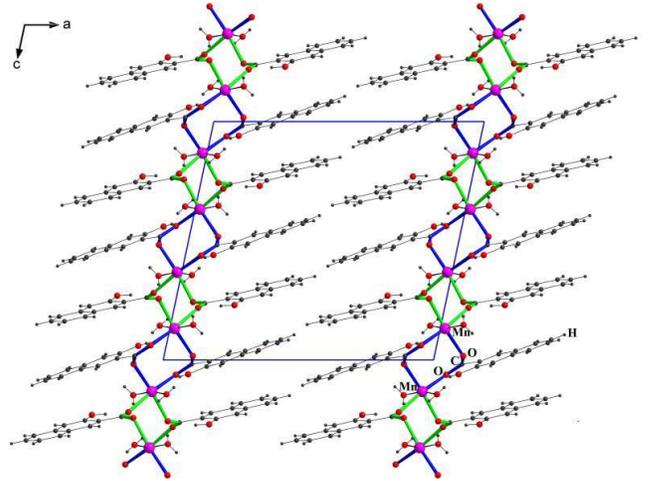,width=85mm}
  \end{center}
  \vspace{-4mm}
  \caption{%
    View along the $b$-axis of the metal-organic quantum magnet 
    $\mathrm{Mn}[\mathrm{C}_{10}\mathrm{H}_{6}(\mathrm{OH})(\mathrm{COO})]_{2}
  \!\times\! 2\mathrm{H}_{2}\mathrm{O}$.
   Bold lines show exchange paths 
   $\mathrm{Mn}-\mathrm{O}-\mathrm{C}-\mathrm{O}-\mathrm{Mn}$.
   The unit cell, denoted by the parallelogram, 
   contains four crystallographically equivalent $\mathrm{Mn}^{2+}$ ions.
  }
  \label{fig:BONS-b}
\end{figure}

\begin{figure}[tb]
  \begin{center}
    \epsfig{file=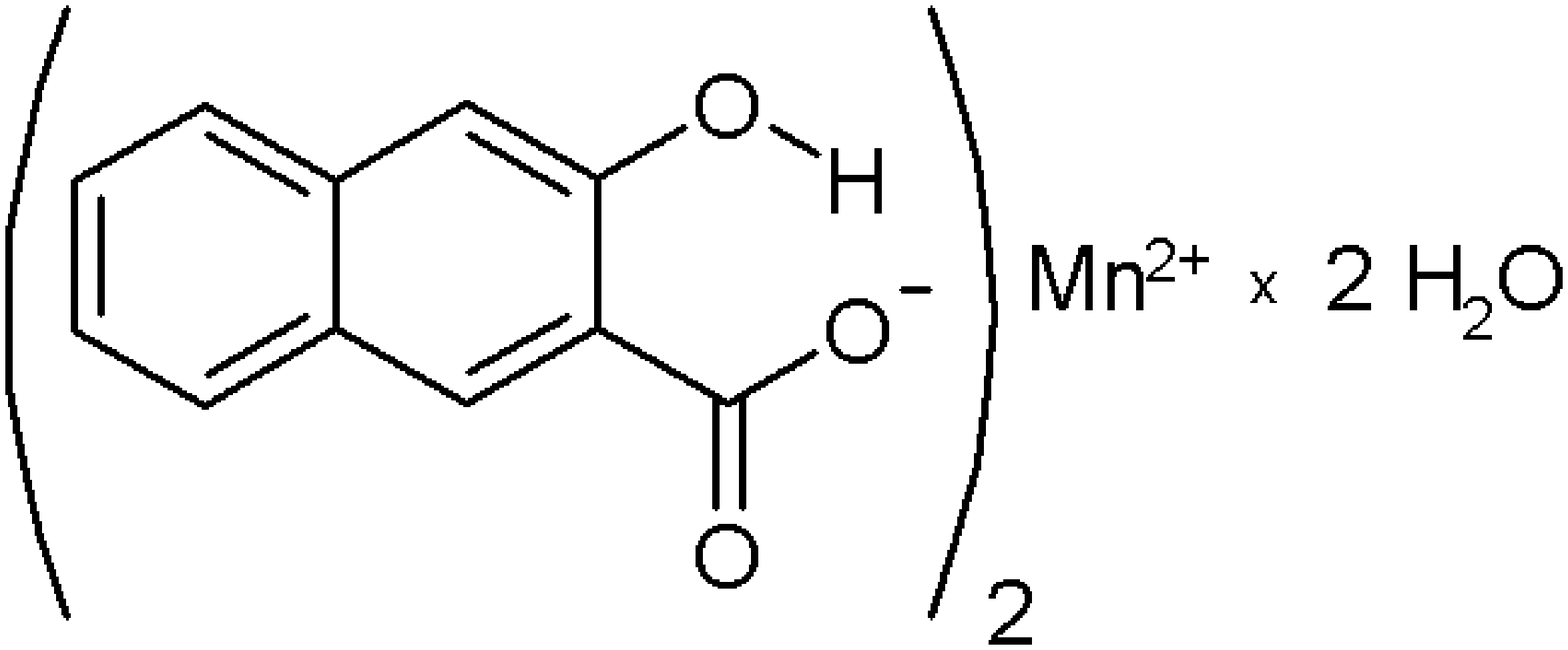,width=85mm}
  \end{center}
  \vspace{-4mm}
  \caption{%
    Chemical formula of 
    $\mathrm{Mn}[\mathrm{C}_{10}\mathrm{H}_{6}(\mathrm{OH})(\mathrm{COO})]_{2}
  \!\times\! 2\mathrm{H}_{2}\mathrm{O}$.
  }
  \label{fig:BONS-chem}
\end{figure}

\begin{figure}[tb]
  \begin{center}
    \epsfig{file=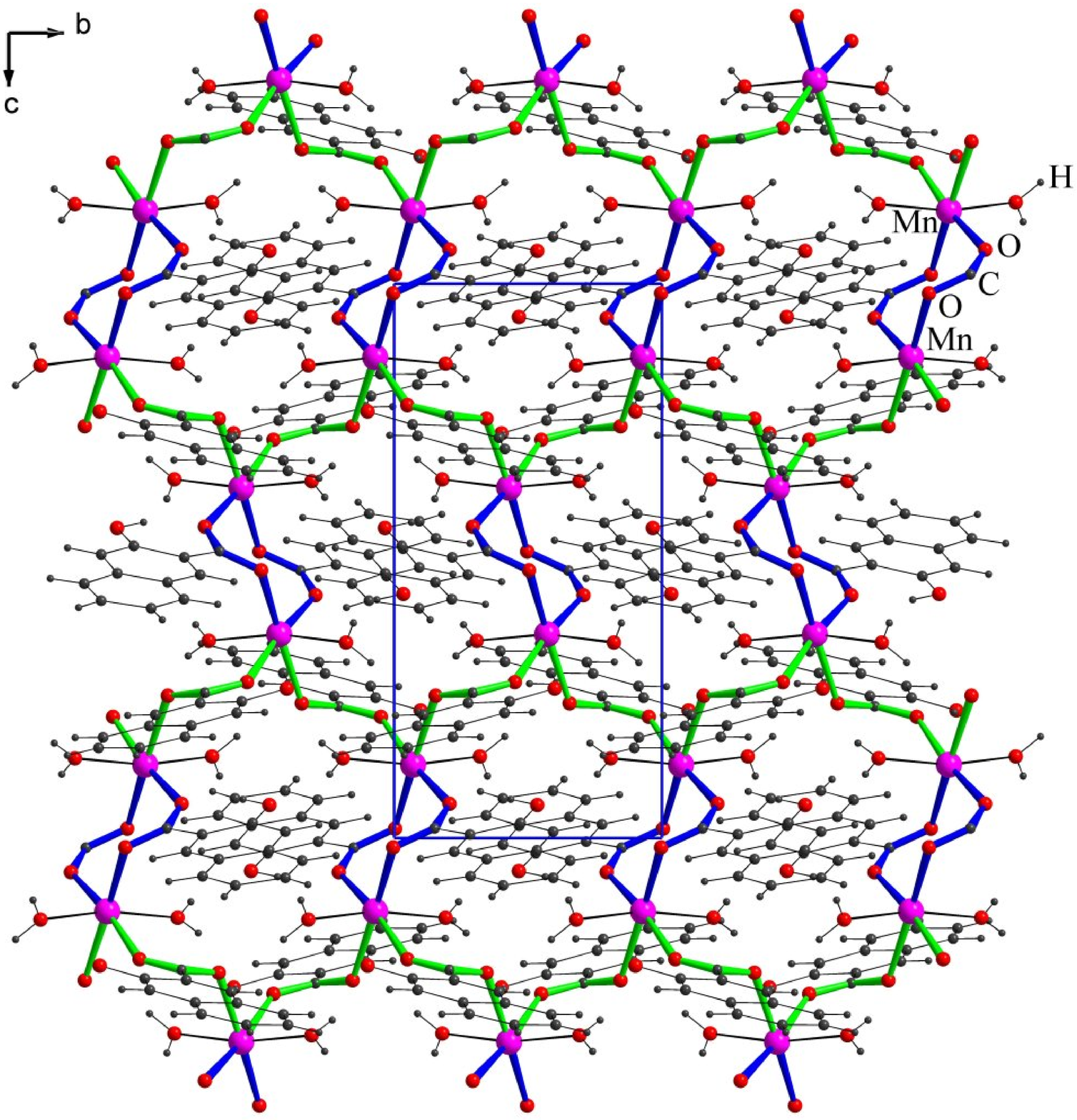,width=85mm}
  \end{center}
  \vspace{-4mm}
  \caption{%
    View on the $(bc)$ plane of the metal-organic quantum magnet 
    $\mathrm{Mn}[\mathrm{C}_{10}\mathrm{H}_{6}(\mathrm{OH})(\mathrm{COO})]_{2}
  \!\times\! 2\mathrm{H}_{2}\mathrm{O}$.
  }
  \label{fig:BONS}
\end{figure}

These effects are of particular importance in low-dimensional magnets, 
e.g.~$2\mathrm{D}$ layer structures with strong magnetic couplings within the layers and 
weak interactions between the layers. Such layer structures can be built up chemically 
from spin-bearing metal ions, which are connected by short bridges, being separated 
by organic fragments of considerable size, see Fig.~\ref{fig:BONS-b}.
Motivated by these considerations we synthesized transition metal complexes 
of o-hydroxy-naphthoic acids. The crystal structure of 
$\mathrm{Mn}[\mathrm{C}_{10}\mathrm{H}_{6}(\mathrm{OH})(\mathrm{COO})]_{2}
  \!\times\! 2\mathrm{H}_{2}\mathrm{O}$ (systematic name: manganese(II) 
3-hydroxy-2-naphthoate dihydrate, Fig.~\ref{fig:BONS-chem}) is of particular interest, 
because the $\mathrm{Mn}^{2+}$ ions form a distorted honeycomb lattice (Fig.~\ref{fig:BONS}).
For the redetermination of the crystal structure, pale brown crystals were slowly 
grown by diffusion of an aqueous solution of 
$\mathrm{Na}[\mathrm{C}_{10}\mathrm{H}_{6}(\mathrm{OH})(\mathrm{COO})]$ 
into an aqueous $\mathrm{MnSO}_{4}$ solution with a buffer layer of water. 
The single crystal X-ray analysis confirmed the previously determined 
structure\cite{Schmidt04} with a higher precision. 
The compound crystallizes in the monoclinic space group  $P2_1/c$ with the lattice 
parameters $a = 17.191(4)\,\mathrm{\AA}$, $b = 7.3448(10)\,\mathrm{\AA}$, 
$c = 15.5279(17)\,\mathrm{\AA}$, $\beta = 101.964(8)^{\circ}$, 
$V = 1918.1(5)\,\mathrm{\AA}^3$.\cite{CCDC} 
The unit cell contains four crystallographically equivalent $\mathrm{Mn}^{2+}$ ions. 

The coupling layer, parallel to the $(bc)$ plane, contains the $\mathrm{Mn}^{2+}$ 
ions, the $\mathrm{COO}^{-}$ and $\mathrm{OH}$ groups as well as water molecules. 
The isolating layer, having a thickness of about $12\,\mathrm{\AA}$ consists of 
the organic naphthalene moieties. 
These naphthalene moieties are only bound together by 
van der Waals contacts between C and H atoms. The relative weakness of these 
interactions is reflected by the morphology of the crystals: the crystals 
grow in ($b$) and ($c$) direction much faster than in ($a$) direction, 
thus forming thin plates parallel to the ($bc$) plane.

The magnetism is due to the $S = 5/2$ manganese ions which form a distorted 
honeycomb pattern parallel to the ($bc$) planes. 
Neighboring ions are connected by carboxylic 
groups, which provide an 
$\mathrm{Mn}-\mathrm{O}-\mathrm{C}-\mathrm{O}-\mathrm{Mn}$ 
magnetic exchange path. There are two 
different exchange paths: the first path contains a single 
$\mathrm{O} - \mathrm{C} - \mathrm{O}$ unit, displayed 
in \jonecolor in Fig.~\ref{fig:BONS}. In the second path 
(marked with \jtwocolor color) the $\mathrm{Mn}^{2+}$ ions are connected by two 
$\mathrm{O} - \mathrm{C} - \mathrm{O}$
moieties simultaneously. The honeycomb layers are well separated from each other; 
the closest distances between $\mathrm{Mn}^{2+}$ ions of different layers 
are as large as $16.282\,\mathrm{\AA}$.

\begin{figure}[t]
  \centering
  \psfrag{J1}{$J_1$}
  \psfrag{J2}{$J_2$}
  \psfrag{J3}{$J_3$}
  \psfrag{d1}{$\boldsymbol{\delta}_1$}
  \psfrag{d2}{$\boldsymbol{\delta}_2$}
  \psfrag{d3}{$\boldsymbol{\delta}_3$}
  \psfrag{a1}{$a_1$}
  \psfrag{a2}{$a_2$}
  \psfrag{a3}{$a_3$}
  \psfrag{f}{$\varphi$}
  \psfrag{x}{$x$}
  \psfrag{y}{$y$}
  \psfrag{sublattice A}{sublattice A}
  \psfrag{sublattice B}{sublattice B}
  \epsfig{file=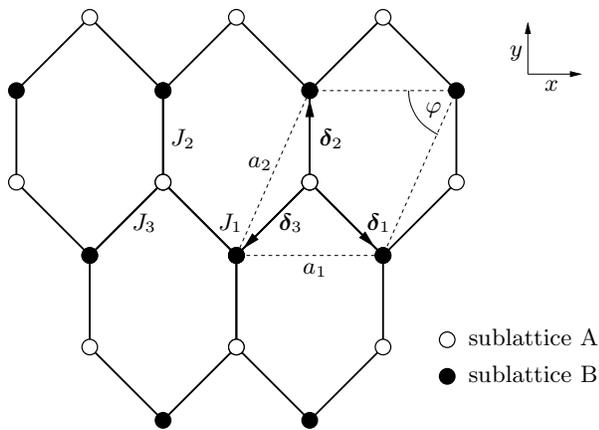,width=220pt}
  \caption{Distorted honeycomb lattice.
    The interactions between spins are displayed as solid lines.  The
    underlying magnetic sublattice is a Bravais lattice and its
    primitive cell can be chosen to be the dashed parallelogram.  The
    corresponding primitive vectors are $ {\bd a}_1 =
    a_{1}{\hat{\bd e}}_x$ and ${\bd a}_2 =
    a_{2}\cos{\varphi}\,{\hat{\bd e}}_x + 
    a_{2}\sin{\varphi}\,{\hat{\bd e}}_y$.  }
  \label{fig:lattice}
\end{figure}

The structure in Fig.~\ref{fig:BONS} suggests that the magnetic
properties of the material can be modeled by a spin $S=5/2$ Heisenberg
magnet on the distorted honeycomb lattice shown in
Fig.~\ref{fig:lattice}. The exchange integrals $J_{\nu} =
J ( {\bd{r}}_i , {\bd{r}}_i + \bd{\delta}_{\nu} )$, $\nu = 1,2,3$,
couple the spin at a given site ${\bd r}_i$ to its nearest neighbors at
${\bd r}_i + {\bd\delta}_{\nu}$.  All exchange integrals $J_\nu$ turn
out to be positive, and 
$\left|\boldsymbol{\delta}_1\right| =
\left|\boldsymbol{\delta}_3\right| \equiv \delta_1 = 5.131(4)\,\mathrm{\AA}$ 
and $J_1 = J_3$, due to the crystal symmetry.
A closer look at the crystal structure in Fig.~\ref{fig:BONS} and a
comparison with the distorted honeycomb lattice in
Fig.~\ref{fig:lattice} reveals that $J_2$ acts along two exchange
paths while $J_1$ results from a single exchange path. 
Therefore we expect $J_2$ to be roughly twice as large as $J_1$.
Furthermore, the honeycomb lattice is bipartite, i.e., it can be
divided into two sublattices, labeled A and B, such that the nearest
neighbors of all sites belonging to sublattice A are located on
sublattice B.  Thus, for positive $J_\nu$ the system is not
frustrated, and when quantum fluctuations are neglected the ground
state shows classical antiferromagnetic N\'eel order. More generally, we
expect long-range antiferromagnetic order to persist in the quantum
mechanical ground state. Therefore, it should be possible to calculate
the magnetic properties of the system within the usual spin-wave
expansion, at least for temperature $T=0$.
Note that the actual structure shown in Fig.~\ref{fig:BONS} has an additional 
distortion in the $x$-direction, resulting in a primitive cell with doubled volume.
Due to the low symmetry of the lattice the 
Dzyaloshinskii-Moriya interaction might play an important role. 
However, we expect the corresponding energy scale to be small in
comparison with $J_1$ and $J_2$, so that in the first approximation
we can neglect this effect.
In the following we therefore always work with the magnetically equivalent 
Bravais lattice shown in Fig.~\ref{fig:lattice}.

Measurements of the magnetization $M(H,T)$ in a magnetic field
$H$ are performed at finite temperatures $T$, where long-range
antiferromagnetic order is ruled out by the Hohenberg-Mermin-Wagner
theorem.\cite{Hohenberg67}  In this case the theoretical justification
for the spin-wave expansion in two dimensions is more subtle.  As long
as there is long-range antiferromagnetic order at $T=0$, it is
reasonable to expect that the low-energy and long-wavelength physics
is still dominated by renormalized spin-waves.  The magnetic
properties of square lattice antiferromagnets in the absence of
uniform external fields have been thoroughly investigated in a
classical work by Chakravarty, Halperin, and Nelson.
\cite{Chakravarty88} Less is known about the low-energy physics of
two-dimensional quantum antiferromagnets subject to a uniform external
magnetic field. 
The external field breaks the rotational symmetry of the Heisenberg
antiferromagnet to O(2), similar to the effect of an XY anisotropy
in the XXZ model.\cite{Fukumoto96} However, the classical ground states
of the two models differ substantially: whereas the XXZ model has a 
collinear ground state, an uniform magnetic field in a Heisenberg 
antiferromagnet leads to a canted classical spin configuration shown in
Fig.~\ref{fig:spingroundstate}.
The zero-temperature magnetization curve $M(H,0)$ of
the square lattice antiferromagnet has been calculated a few years ago
by Zhitomirsky and Nikuni~\cite{Zhitomirsky98} within the spin-wave
expansion.  For finite temperatures, $M(H,T)$ has been extrapolated
from numerical diagonalizations of finite clusters.\cite{Fabricius92}
We are not aware of any analytical calculations in the literature
of $M(H,T)$ for two-dimensional quantum Heisenberg antiferromagnets at $T > 0$.
In this work, we calculate $M(H,T)$ using a modified spin-wave
approach \cite{Kollar03} which takes the absence of a spontaneous
staggered magnetization at finite temperatures into account.  
Our theoretical results for the magnetization curves as well as for the
zero-field susceptibility $\chi(T)$ show a satisfactory agreement with
our measurements for the compound
$\mathrm{Mn}[\mathrm{C}_{10}\mathrm{H}_{6}(\mathrm{OH})(\mathrm{COO})]_{2}
\!\times\! 2\mathrm{H}_{2}\mathrm{O}$.

The rest of the paper is organized as follows. In
Sec.~\ref{sec:general} we review the formalism of the spin-wave
expansion for non-collinear spin configurations. In
Sec.~\ref{sec:honeycomb} this method is applied to an antiferromagnet on a
bipartite lattice in the presence of a uniform magnetic field.
Expressions for the magnetization, the staggered magnetization and the
uniform susceptibility for the material of interest are obtained. We explain how a
self-consistently determined staggered field is used to regularize
divergencies at finite temperature. In Sec.~\ref{sec:results} we
present results and compare with experimental measurements.
Finally, in Sec.~\ref{sec:discussion} we present our conclusion.

\section{Spin waves in non-collinear spin configurations}
\label{sec:general}

\begin{figure}[tb]
  \centering
  \psfrag{x}{$x$}
  \psfrag{z}{$z$}
  \psfrag{t}{$\theta$}
  \psfrag{ma}{${\hat{\bd m}}_{\mathrm{A}}$}
  \psfrag{mb}{${\hat{\bd m}}_{\mathrm{B}}$}
  \psfrag{m}{$\langle {\bd S}_i \rangle $}
  \psfrag{B}{$ B {\hat{\bd e}}_x $}
  \psfrag{Bs}{$\zeta_i B_s {\hat{\bd e}}_z$}
  \psfrag{e2a}{$ {\hat{\bd e}}_\mathrm{A}^2$}
  \psfrag{e2b}{$ {\hat{\bd e}}_B^2$}
  \epsfig{file=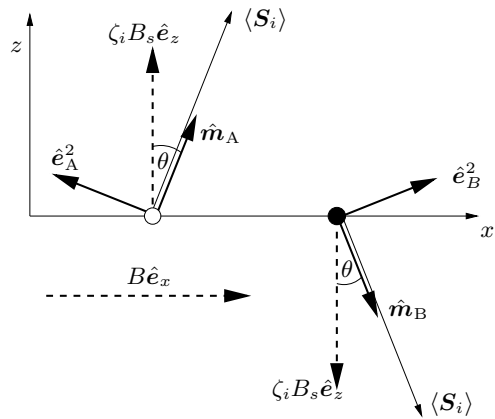,width=65mm}
  \vspace{-4mm}
  \caption{%
    Spin configuration $\langle \bd{S}_i \rangle$ in the classical ground state
    of a two-sublattice antiferromagnet .  The dashed arrows
    represent a uniform magnetic field $ B \hat{\bd{e}}_x$ in the
    $x$-direction and a staggered magnetic field $ \zeta_i B_s
    \hat{\bd{e}}_z$ in the $z$-direction.  The small solid arrows
    represent the vectors of a ``co-moving'' basis that matches the
    direction defined by the local magnetization $\langle \bd{S}_i
    \rangle$.  Not shown are the basis vectors $\hat{\bd{e}}_\mathrm{A}^1 =
    \hat{\bd{e}}_\mathrm{B}^1 = \hat{\bd{e}}_y$ which point into the plane of
    the paper.  }
  \label{fig:spingroundstate}
\end{figure}

In the presence of a homogeneous magnetic field an antiferromagnet on
a bipartite lattice has a non-collinear, canted spin configuration as
shown in Fig.~\ref{fig:spingroundstate}. 
We choose a coordinate system such that the uniform external field
points along the $x$-axis, and the staggered magnetization is directed
along the $z$-axis.  The low temperature physics is dominated by
spin-wave excitations.  To obtain their spectrum we should quantize
the spin-operators in a spatially-dependent (``co-moving'') coordinate
system that matches for each site the axis defined by the expectation
value $\langle \bd{S}_i \rangle$ of the spin operator.

More generally, the problem of calculating the spin excitations of a
Heisenberg magnet subject to an arbitrary inhomogeneous magnetic field
${\bd{B}}_i$ can be formulated and solved in a coordinate-free vector
notation.\cite{Schuetz03}  Consider the general Heisenberg hamiltonian
\begin{equation}
  \hat{H} = \frac{1}{2} \sum_{  i,j}   J_{ij} {\bd{S}}_i \cdot
  {\bd{S}}_j   - g \mu_{\text{B}} \sum_{i  }  {\bd{B}}_i \cdot
  {\bd{S}}_i
  \,,
  \label{eq:Hamiltonian}
\end{equation} 
where $J_{ij} = J ( \bd{r}_i, \bd{r}_j ) $ are some arbitrary exchange
couplings, the sums are over all sites ${\bd{r}}_i$ of a
$D$-dimensional lattice consisting of $N$ sites, and the ${\bd{S}}_i$
are spin-$S$ operators normalized such that ${\bd{S}}_i^2 = S ( S+1
)$.  The last term represents the Zeeman energy, where $g$ is the
gyromagnetic factor and $\mu_{\mathrm{B}}$ is the Bohr magneton.  We assume that
the external magnetic field is sufficiently strong to induce permanent
magnetic dipole moments ${\bd{m}}_i = g \mu_{\mathrm{B}} \langle {\bd{S}}_i
\rangle$, where $\langle \ldots \rangle$ denotes the thermal
equilibrium average.  It is then convenient to decompose the spin
operators as ${\bd{S}}_i = S^{\parallel}_i \hat{\bd{m}}_i +
{\bd{S}}^{\bot}_i$, where ${\bd{S}}_i^{\bot} \cdot \hat{\bd{m}}_i =
0$, and $\hat{\bd{m}}_i = \bd{m}_i / | {\bd{m}}_i |$ is a unit vector
in the direction of $\bd{m}_i$.  Substituting this decomposition into
Eq.~(\ref{eq:Hamiltonian}) we obtain $\hat{H} = \hat{H}^{\parallel} +
\hat{H}^{\bot} + \hat{H}^{\prime}$, with
\begin{eqnarray}
  \hat{H}^{\parallel}\,\, & = & \frac{1}{2} \sum_{  i,j}  J_{ij}  
  \hat{\bd{m}}_i \cdot \hat{\bd{m}}_j
  {{S}}_i^{ \parallel} 
  {{S}}_j^{\parallel}   - \sum_{i  }  {\bd{h}}_i  \cdot \hat{\bd{m}}_i 
  {{S}}_i^{\parallel}
  \,,
  \label{eq:Hparallel}
  \\
  \hat{H}^{\bot} & = & \frac{1}{2} \sum_{  i,j}  J_{ij}  
  {\bd{S}}^{\bot}_i \cdot {\bd{S}}^{\bot}_j  
  \,,
  \label{eq:Hbot}
  \\
  \hat{H}^{\prime}\,\, & = &  - \sum_{  i}   {\bd{S}}^{\bot}_i
  \cdot \bigl(  {\bd{h}}_i -
  \sum_j J_{ij}  
  {{S}}_j^{ \parallel}   \hat{\bd{m}}_j   \bigr)
  \,,
  \label{eq:Hrest}
\end{eqnarray}
where ${\bd{h}}_i = g \mu_{\mathrm{B}} {\bd{B}}_i $.  Note that
$\hat{H}^{\prime}$ describes the coupling between the transverse and
the longitudinal spin fluctuations.  The classical ground state energy
${E}_0^{\rm cl}$ is obtained by replacing $S_i^{\parallel} \rightarrow
S$ in Eq.~(\ref{eq:Hparallel}) and by finding the configuration $\{
\hat{\bd{m}}_i \}$ that minimizes the resulting classical hamiltonian
\begin{equation}
  H^{\rm cl} =
  \frac{S^2}{2} \sum_{  i,j}  J_{ij}  
  \hat{\bd{m}}_i \cdot \hat{\bd{m}}_j
  - S \sum_{i  }  {\bd{h}}_i  \cdot \hat{\bd{m}}_i 
  \label{eq:Hcl}
  \; .
\end{equation}
A necessary condition for an extremum of Eq.~(\ref{eq:Hcl}), taking
into account the constraints $\hat{\bd{m}}_i^2 =1$,
is~\cite{Schuetz03}
\begin{equation}
  \hat{\bd{m}}_i
  \times
  \Big(
  {\bd{h}}_i - S \sum_j J_{ij}  \hat{\bd{m}}_j
  \Big)
  = 0
  \,. 
  \label{eq:classical}
\end{equation} 
For given ${\bd{h}}_i $ and $J_{ij}$, this is a system of non-linear
equations for the spin directions $\hat{\bd{m}}_i$ in the classical
ground state. Using Eq.~(\ref{eq:classical}), the part
$\hat{H}^{\prime}$ of the hamiltonian describing the coupling between
transverse and longitudinal fluctuations can be written as
\begin{equation}
  \hat{H}^{\prime} = - \sum_{ i,j} J_{ij} 
  ( {\bd{S}}_i^{\bot} \cdot \hat{\bd{m}}_j ) ( S -S_j^{\parallel} )
  \; .
  \label{eq:Hprime}
\end{equation}
Let us expand the transverse components of $ {\bd{S}}^{\bot}_i $ in a
spherical basis, $ {\bd{S}}^{\bot}_i $ $=$ $\frac{1}{2} \sum_{ p = \pm
} S_i^{-p} {\bd{e}}^{p}_i $, with the spherical basis vectors
${\bd{e}}^{p}_i $ $=$ $ \hat{\bd{e}}^1_i + i p \hat{\bd{e}}^2_i$, $p $
$=$ $ \pm$, where $ \left\{ \hat{\bd{e}}_i^{1}, \hat{\bd{e}}_i^2 ,
  \hat{\bd{m}}_i \right\}$ is a local orthogonal triad of unit
vectors. The transverse part of our spin hamiltonian can then be
written as
\begin{equation}
  \hat{H}^{\bot}  =  \frac{1}{8} \sum_{  i,j}  \sum_{ p, p^{\prime} } J_{ij}  
  ( {\bd{e}}^{p}_i \cdot {\bd{e}}^{p^{\prime}}_j )
  {{S}}^{-p}_i 
  {{S}}^{-p^{\prime}}_j  
  \,.
  \label{eq:Hbot2}
\end{equation}
The basis vectors $ \hat{\bd{e}}_i^{1}, \hat{\bd{e}}_i^2 $ are not
unique: any rotation around $\hat{\bd{m}}_i$ yields an equally
acceptable transverse basis.

So far, no approximation has been made. To obtain the spin-wave
spectrum, we expand the spin operators in terms of canonical boson
operators $b_i$ as usual \cite{Dyson56,Maleev57}, $S^{\parallel}_i$ $=$
$S-b_i^{\dagger}b_i^{\phantom{\dagger}}$ and $S^{+}_i$ $=$
$(S^{-}_i)^\dagger$ $=$
$\sqrt{2S}\,b_i^{\phantom{\dagger}}\,(1+O(S^{-1}))$.  Within the
linear spin-wave approximation the hamiltonian becomes $\hat{H}
\approx E_0^{\rm cl} + \hat{H}_2$, where $E_0^{\rm cl}$ is the minimum
of the classical hamiltonian $H^{\rm cl}$ in Eq.~(\ref{eq:Hcl}), and
\begin{eqnarray}
  \hat{H}_2 & = & \frac{S}{4} \sum_{i,j} J_{ij} [
  (\bd{e}^{+}_i \cdot \bd{e}^-_j ) \; b_i^{\dagger} b_j
  + (\bd{e}^{-}_i \cdot \bd{e}^+_j ) \; b_j^{\dagger} b_i
  \nonumber 
  \\
  & & \hspace{12mm} +
  (\bd{e}^{+}_i \cdot \bd{e}^+_j ) \; b_i^{\dagger} b_j^{\dagger}
  + (\bd{e}^{-}_i \cdot \bd{e}^-_j ) \; b_j b_i
  ]
  \nonumber
  \\
  & - &  \frac{S}{2} \sum_{i,j} J_{ij}  ( {\hat{\bd{m}}}_i \cdot \hat{\bd{m}}_j )
  [ b^{\dagger}_i b_i + b^{\dagger}_j b_j ]
  \nonumber
  \\
  & + &\sum_i ( \bd{h}_i \cdot \hat{\bd{m}}_i)  b^{\dagger}_i b_i
  \label{eq:H2}
  \; .
\end{eqnarray} 
Note that the contribution from $\hat{H}^{\prime}$ in
Eq.~(\ref{eq:Hprime}) is of order $S^{1/2}$ and hence can be neglected
within linear spin-wave theory.  Eq.~(\ref{eq:H2}) together with
Eqs.~(\ref{eq:Hcl}) and (\ref{eq:classical}) completely determine the
spin-wave spectrum of any Heisenberg magnet in an arbitrary
inhomogeneous field.

\section{Spin waves in the distorted honeycomb lattice}
\label{sec:honeycomb}

\subsection{Classical ground state}

Let us apply the general formalism of the previous section to our
bipartite lattice antiferromagnet in a uniform external magnetic field $B
\hat{\bd{e}}_x$ along the $x$ axis. We denote by $\hat{\bd{e}}_\alpha$
fixed unit vectors in direction $\alpha=x,y,z$.  For technical reasons 
we introduce an additional staggered magnetic field $\zeta_i B_s
\hat{\bd{e}}_z$ in the $z$-direction, where $\zeta_i = 1$ if
${\bd{r}}_i$ belongs to the A-sublattice and $\zeta_i = -1$ if
${\bd{r}}_i$ belongs to the B-sublattice.  
This auxiliary staggered field will be determined self-consistently 
in Sec.~\ref{subsec:magn}  
to insure a vanishing staggered magnetization at finite
temperatures. The total magnetic field is thus
\begin{equation}
  \bd{h}_i = g \mu_{\mathrm{B}} [B \hat{\bd{e}}_x
  +  \zeta_i B_s \hat{\bd{e}}_z]
  \; .
  \label{eq:hidef}
\end{equation}
The classical ground state configuration is then $\hat{\bd{m}}_i =
\zeta_i \cos \theta \hat{\bd{e}}_z + \sin \theta \hat{\bd{e}}_x $, as
shown in Fig.~\ref{fig:spingroundstate}.

For convenience we introduce the notation $n_0 = \cos \theta $ and
$m_0 = \sin \theta$.  Physically, $m_0$ corresponds to the classical
limit ($S \rightarrow \infty$) of the normalized uniform magnetization
\begin{equation}
  m  =  \frac{1}{NS} \sum_{i} \langle \hat{\bd{e}}_x \cdot \bd{S}_i \rangle
  \; , 
  \label{eq:mdef}
\end{equation}
while $n_0$ corresponds to the $S \rightarrow \infty$ limit of the
normalized staggered magnetization
\begin{equation}
  n  =  \frac{1}{NS} \sum_{i} \zeta_i \langle \hat{\bd{e}}_z \cdot 
  \bd{S}_i \rangle
  \; .
  \label{eq:ndef}
\end{equation}
By symmetry, the uniform magnetization points into the $x$-direction,
while the staggered magnetization points into the $z$-direction.  The
natural dimensionless measure for the strength of the fields is ${h} =
\chi_0 g \mu_{\mathrm{B}} B$ and ${h}_s = \chi_0 g \mu_{\mathrm{B}} B_s$, 
where $\chi_0 = (2 \tilde{J}_0 S )^{-1}$ 
is the classical uniform susceptibility.  Here
$\tilde{J}_0 = \sum_{\nu} J_{\nu}$ is the ${\bd{k}} =0$ component of
the Fourier transform of the exchange couplings
\begin{equation}
  \tilde{J}_{\bd{k}} = \sum_{\nu} e^{ -i {\bd{k}}
  \cdot \bd{\delta}_{\nu} } J_{\nu}\,.
  \label{eq:jdef} 
\end{equation}
For the special choice of the field $\bd{h}_i$ given in
Eq.~(\ref{eq:hidef}) our general Eq.~(\ref{eq:classical}) reduces 
to the simple relation
\begin{equation}
  {h} = m_0 [ 1 + {h}_s / n_0 ]
  \; ,
  \label{eq:hh}
\end{equation}
which together with $n_0^2 + m_0^2 =1$ determines the classical N\'eel
order parameter $n_0$ and the classical uniform magnetization $m_0$ as
functions of the fields $h$ and $h_s$.  Note that $\hat{\bd{m}}_\mathrm{A}
\cdot \hat{\bd{m}}_\mathrm{B} = m_0^2 - n_0^2$, and with the special transverse
basis shown in Fig.~\ref{fig:spingroundstate}
\begin{eqnarray}
  \bd{e}_\mathrm{A}^p \cdot \bd{e}_\mathrm{B}^{p^{\prime}} & = & 2 [\delta_{ p,p^{\prime}} n_0^2
  +  \delta_{p, - p^{\prime}} m_0^2 ] \; ,
  \label{eq:epep}
  \\
  \hat{\bd{m}}_\mathrm{A} \cdot {\bd{e}}_\mathrm{B}^p & = &
  2 i p n_0 m_0 =
  - \hat{\bd{m}}_\mathrm{B} \cdot {\bd{e}}_\mathrm{A}^p
  \; \; .
  \label{eq:medot}
\end{eqnarray}

\subsection{Spin-wave dispersion}

\begin{figure}[tb]
  \centering
  \psfrag{x}{$k_x$}
  \psfrag{y}{$k_y$}
  \psfrag{b1}{${\bd b}_{1}$}
  \psfrag{b2}{${\bd b}_{2}$}
  \psfrag{f}{$\varphi$}
  \epsfig{file=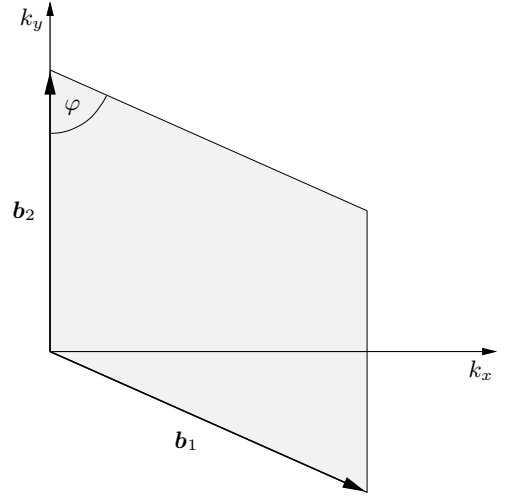,width=65mm}
  \vspace{-2mm}
  \caption{%
    Reduced Brillouin zone of the distorted honeycomb lattice.  The
    primitive vectors are ${\bd b}_1 =
    \frac{2\pi}{a_1\sin{\varphi}} (\sin{\varphi}\,{\hat{\bd e}}_x -
    \cos{\varphi}\,{\hat{\bd e}}_y) $ and ${\bd b}_2 =
    \frac{2\pi}{a_2\sin{\varphi}}{\hat{\bd e}}_y$, where $a_1$,
    $a_2$ and the angle $\varphi$ are defined in
    Fig.~\ref{fig:lattice}.  }
  \label{fig:BZ}
\end{figure}

To obtain the spin-wave dispersion, we must diagonalize $\hat{H}_2$ in
Eq.~(\ref{eq:H2}) for the special ground-state spin configuration
discussed above.  Therefore, we first perform Fourier transformations
separately on each sublattice,
\begin{subequations}
  \begin{eqnarray}
    b_i & = & \sqrt{ \frac{2}{N} } \sum_{\bd{k}} e^{ i {\bd{k}} \cdot {\bd{r}}_i }
    a_{ \bd{k}} \; \; , \; \; \mbox{for $\bd{r}_i \in$ A }
    \; ,
    \label{eq:AFT}
    \\
    b_i & = & \sqrt{ \frac{2}{N} } \sum_{\bd{k}} e^{ i {\bd{k}} \cdot {\bd{r}}_i }
    b_{ \bd{k}} \; \; , \; \; \mbox{for $\bd{r}_i \in$ B }
    \; ,
    \label{eq:BFT}
  \end{eqnarray}
\end{subequations}
where the wave-vector sums are over the reduced (magnetic) Brillouin
zone of the honeycomb lattice shown in Fig.~\ref{fig:BZ}.
With the above definitions we obtain
\begin{eqnarray}
  \hat{H}_2 & = & \tilde{J}_0 S \sum_{ \bd{k}} \bigl[
  A ( a^{\dagger}_{\bd{k}} a^{}_{\bd{k}} + b^{\dagger}_{\bd{k}} b^{}_{\bd{k}} )
  + B^{}_{\bd{k}} b^{}_{ - \bd{k}} a^{}_{\bd{k}}  + B_{\bd{k}}^{\ast} 
  a^{\dagger}_{ \bd{k}} b^{\dagger}_{- \bd{k}}   
  \nonumber
  \\
  &  & \hspace{15mm} +
  C^{}_{ \bd{k}}  b^{\dagger}_{  \bd{k}} a^{}_{\bd{k}} + C_{\bd{k}}^{\ast} 
  a^{\dagger}_{\bd{k}} b^{}_{ \bd{k}}
  \bigr]
  \; ,
  \label{eq:H2a}
\end{eqnarray}
where $A =1 + 2 {h}_s / n_0$, $B_{\bd{k}} = n_0^2 \tilde{J}_{\bd{k}} /
\tilde{J}_0$, and $C_{\bd{k}} = m_0^2 \tilde{J}_{\bd{k}} /
\tilde{J}_0$.  On a honeycomb lattice $\tilde{J}_{\bd{k}} =|
\tilde{J}_{\bf{k}} | e^{i \phi_{\bd{k}} }$ is complex, so that
$B_{\bd{k}} = | B_{\bf{k}} | e^{i \phi_{\bd{k}} }$ and $C_{\bd{k}} = |
C_{\bf{k}} | e^{i \phi_{\bd{k}} }$.  Using $\phi_{ - \bd{k}} = -
\phi_{ \bd{k}}$, it is easy to see that these phase factors can be
removed from Eq.~(\ref{eq:H2a}) via the gauge transformation
$\tilde{a}_{ \bd{k}} = e^{i \phi_{\bd{k}}} a_{\bd{k}}$.  Introducing
then new canonical boson operators
\begin{equation}
  c_{\bd{k} \sigma } = \frac{1}{\sqrt{2}} \left[ \tilde{a}_{\bd{k}} + \sigma
    b_{\bd{k}} \right]
  \;  \; , \; \; \sigma = \pm 1
  \; ,
  \label{eq:cpmdef}
\end{equation}
the hamiltonian (\ref{eq:H2a}) assumes the block-diagonal form,
\begin{eqnarray}
  \hat{H}_2 & = & \frac{\tilde{J}_0 S}{2} \sum_{ \bd{k}  \sigma} \bigl[
  (A + \sigma | C^{}_{\bd{k}} | ) ( c^{\dagger}_{\bd{k} \sigma} c^{}_{\bd{k} \sigma} 
  + c^{\dagger}_{-\bd{k} \sigma} c^{}_{-\bd{k} \sigma} )
  \nonumber
  \\
  &  & \hspace{12mm} +
  \sigma | B^{}_{ \bd{k}} |  
  ( c^{\dagger}_{  \bd{k} \sigma} c^{\dagger }_{-\bd{k} \sigma} +  
  c^{}_{ \bd{k} \sigma} c^{}_{ -\bd{k} \sigma} )
  \bigr]
  \; .
  \label{eq:H2b}
\end{eqnarray}
The diagonalization is completed by means of the Bo\-go\-liu\-bov
transformation,
\begin{equation}
  \left( \begin{array}{c}
      c_{ \bd{k} \sigma } \\
      c^{\dagger}_{ - \bd{k} \sigma }  \end{array}
  \right) =
  \left( \begin{array}{cc}
      u_{ \bd{k} \sigma} & - \sigma v_{\bd{k} \sigma} \\
      - \sigma  v_{\bd{k} \sigma} & u_{ \bd{k} \sigma} \end{array} \right)
  \left( \begin{array}{c}
      d_{ \bd{k} \sigma } \\
      d^{\dagger}_{ - \bd{k} \sigma }  \end{array}
  \right)
  \; ,
  \label{eq:bogoliubov}
\end{equation}
where
\begin{subequations}
  \begin{eqnarray}
    u_{ \bd{k} \sigma } & = & 
    \sqrt{ \frac{ A + \sigma | C_{\bd{k}} | + \epsilon_{\bd{k} \sigma} }{ 2 \epsilon_{\bd{k} \sigma}} }
    \; ,
    \label{eq:ukdef}
    \\
    v_{ \bd{k} \sigma } & = & 
    \sqrt{ \frac{ A + \sigma | C_{\bd{k}} | - \epsilon_{\bd{k} \sigma} }{ 2 \epsilon_{\bd{k} \sigma}} }
    \; ,
    \label{eq:vkdef}
  \end{eqnarray}
\end{subequations}
with the dimensionless energy dispersion
\begin{equation} 
  \epsilon_{\bd{k} \sigma} = \sqrt{
    (  A + \sigma | C_{\bd{k}} | )^2 - | B_{\bd{k}} |^2 }
  \label{eq:epsilondef}
  \; .
\end{equation}
Defining $\gamma_{\bd{k}} = \tilde{J}_{\bd{k}} / \tilde{J}_0$, we may
write
\begin{equation}
  \epsilon_{\bd{k} \sigma} = 
  \Bigl[
  \bigl( 1 +  \frac{2 {h}_s}{n_0} + \sigma | \gamma_{\bd{k}} | \bigr)
  \bigl( 1 +  \frac{2 {h}_s}{n_0} - 
  \sigma ( n_0^2 - m_0^2 )| \gamma_{\bd{k}} | \bigr) \Bigr]^{1/2}
  \; .
  \label{eq:epsilon2}
\end{equation}
In terms of the new operators $d_{\bd{k} \sigma}$ the quadratic spin
wave hamiltonian $\hat{H}_2$ is diagonal,
\begin{equation}
  \hat{H}_2 = \tilde{J}_0 S \sum_{ \bd{k} \sigma} 
  \left\{ \epsilon^{}_{ \bd{k} \sigma} d^{\dagger}_{ \bd{k} \sigma}
    d^{}_{ \bd{k} \sigma} + \frac{1}{2} \left[ \epsilon_{ \bd{k} \sigma}
      - (A + \sigma | C_{\bd{k}} | ) \right] \right\}
  \; .
\end{equation}
The low temperature properties of the magnet are determined by the long-wavelength
behavior of the spin-wave dispersions, which follow from the expansion
for small ${\bd k}$,
\begin{equation} 
  |\gamma_{\bd k}| \approx 1 - \frac12 \sum_{\alpha\beta} k_{\alpha} A_{\alpha\beta}k_{\beta}\,, 
\end{equation} 
where ${\bd A}$ is a matrix with elements
\begin{equation} 
  A_{\alpha\beta} =  
  \sum_{\nu}  \frac{J_\nu}{\tilde{J}_0}  (  \bd{\delta}_{\nu}  \cdot \hat{\bd{e}}_{\alpha} )
  ( \bd{\delta}_{\nu}  \cdot \hat{\bd{e}}_{\beta} )
  - 
  \sum_{\nu,\nu'} \frac{J_\nu J_{\nu'}}{\tilde{J}_0^2}   
  (  \bd{\delta}_{\nu}  \cdot \hat{\bd{e}}_{\alpha} )
  ( \bd{\delta}_{\nu^{\prime}}  \cdot \hat{\bd{e}}_{\beta} )
  \,. 
  \label{eq:Amatrixdef}
\end{equation} 
Since ${\bd A}$ is symmetric, an orthogonal basis can always be chosen
such that ${\bd A}$ is diagonal, with eigenvalues $A_{\alpha}$. In this basis
\begin{equation} 
  |\gamma_{\bd k}| \approx 1 - \frac12\sum_{\alpha} A_{\alpha} k_{\alpha}^2\,. 
\end{equation} 
The matrix ${\bd A}$ is positive, since 
\begin{equation} 
  |\gamma_{\bd k}| \leq \sum_{\nu} \left|\frac{J_{\nu}}{\tilde{J}_0}\right| = 1\,, 
\end{equation} 
where the last equality assumes that all couplings have the same sign.
We can thus define effective length scales $\ell_\alpha$ by setting
$A_{\alpha} = \ell_{\alpha}^2$.  For a $D$-dimensional hypercubic
lattice with lattice spacing $a$ we have $\ell^2_{\alpha}={a^2}/{D}$.
For our honeycomb lattice shown in Fig.~\ref{fig:lattice} with $ |
\bd{\delta}_1 | = | \bd{\delta} _3 | $ and $J_1 = J_3$ the
eigenvectors of $\bd{A}$ are parallel to the $x$-axis and the
$y$-axis, with corresponding eigenvalues $\ell_x^2 = (J_1 / 2
\tilde{J}_0 ) a_1^2$ and $\ell_y^2 = (2 J_1 J_2 / \tilde{J}_0^2 )
a_2^2 \sin^2 \varphi$.  The spin-wave velocities $c_{\alpha} =
\tilde{J}_0 S \ell_{\alpha}$ along the two principal directions are
thus
\begin{eqnarray}
  c_x &=& S \sqrt{ \frac{J_1 \tilde{J}_0 }{2}}\,a_1\,,\\
  c_y &=& S  \sqrt{ 2 J_1 J_2} \,a_2\sin{\varphi}\,.
\end{eqnarray}
Note that for $ J_2 \rightarrow 0$ the velocity $c_y$ vanishes, so
that the system becomes one-dimensional, as is obvious from
Fig.~\ref{fig:lattice}.  On the other hand, for $J_1 \rightarrow 0$
both velocities vanish, because in this limit the system consists of
decoupled dimers.

For ${h}_s =0$ only the mode $\epsilon_{ \bd{k}-}$ is gapless for
${\bd{k}} \rightarrow 0$, while the mode $\epsilon_{ \bd{k}+}$ has the
gap $2 m_0$.  To give a more explicit form for the long-wavelength
spin-wave dispersions, we further assume $h_s \ll n_0$. Then
\begin{eqnarray} 
  \epsilon_{{\bd k}-} & \approx & n_0 \left[\frac{4h_s}{n_0} 
    +\sum_{\alpha}(\ell_{\alpha} k_{\alpha})^2\right]^{1/2}\,,
  \label{eq:epsilonminuslong}
  \\ 
  \epsilon_{{\bd k}+} & \approx & \Bigl[ 4m_0^2+\frac{4h_s}{n_0}(1+m_0^2) 
  \nonumber
  \\
  & & 
  +(n_0^2-2m_0^2)\sum_{\alpha}(\ell_{\alpha}k_{\alpha})^2 \Bigr]^{1/2}\,.   
  \label{eq:epsilonpluslong}
\end{eqnarray} 
For $n_0\to0$ the expansion (\ref{eq:epsilonminuslong}) is not
appropriate any longer and for $h_s=0$ the dispersion $\epsilon_{{\bd k}-}$
becomes purely quadratic at $n_0=0$.  Before this happens, there is a
critical field $0<h^*<1$ at which the curvature of the dispersion
$\epsilon_{{\bd k}-}$
changes sign. The positive curvature for $h>h^*$ results in an
instability of magnons towards a spontaneous decay into two magnon
states.\cite{Zhitomirsky99} Furthermore, if an anisotropic exchange is
considered, the anisotropy gap $\Delta$ is strongly renormalized by
magnon interactions.\cite{Maleyev00,Syromyatnikov01} As the influence
of these instabilities on the thermodynamic properties is unclear at
the moment, they will not be further considered in this work.

\subsection{Uniform and staggered magnetization}
\label{subsec:magn}

We now calculate the leading spin-wave corrections to the normalized
uniform- and the staggered magnetization as defined in
Eqs.~(\ref{eq:mdef}) and (\ref{eq:ndef}).  A standard expansion in
powers of $1/S$ gives
\begin{eqnarray}
  m & = & \frac{m_0^2}{{h}} \left[ 1 + \frac{n {h}_s}{n_0^2}
    - \frac{ F (   {h} , {h}_s )   }{S}  \right]
  \label{eq:Fsw}
  \; ,
  \\
  n & = & \frac{1}{n_0} \left[ 1 - m_0 m - \frac{  I ( {h} , {h}_s )  }{S} 
  \right]
  \label{eq:Isw}
  \; ,
\end{eqnarray}
where
\begin{equation}
  F (   {h} , {h}_s )  = \frac{1}{N} \sum_{ \bd{k} \sigma}
  \frac{  n_{\bd{k} \sigma}  + \frac{1}{2} }{ \epsilon_{ \bd{k} \sigma} } 
  \sigma | \gamma_{\bd{k}} |   
  \Bigl( 1 + \frac{2 {h}_s}{n_0} + \sigma | \gamma_{\bd{k}} | \Bigr) 
  \; ,
  \label{eq:Fdef}
\end{equation}
and
\begin{equation}
  I (   {h} ,{h}_s )  = - \frac{1}{2} + \frac{1}{N} \sum_{ \bd{k} \sigma}
  \frac{  n_{\bd{k} \sigma}  + \frac{1}{2} }{ \epsilon_{ \bd{k} \sigma} } 
  \Bigl( 1 + \frac{2 {h}_s}{n_0} + \sigma m_0^2 | \gamma_{\bd{k}} | \Bigr)
  \; .
  \label{eq:Idef}
\end{equation}
Here $n_{\bd{k} \sigma} = [ e^{ \tilde{J}_0 S \epsilon_{\bd{k} \sigma}
  /T } -1 ]^{-1}$ is the Bose function.  The parameters $n_0$ and
$m_0$ on the right-hand sides of Eqs.~(\ref{eq:Fsw}--\ref{eq:Idef})
are determined as functions of the fields $h$ and $h_s$ by
Eq.~(\ref{eq:hh}) and $n_0^2 + m_0^2 =1$.  Note that for $S
\rightarrow \infty$ the solutions of Eqs.~(\ref{eq:Fsw}) and
(\ref{eq:Isw}) correctly approach $n =n_0$ and $m = m_0$: in this
limit Eq.~(\ref{eq:Fsw}) reduces to Eq.~(\ref{eq:hh}), while
Eq.~(\ref{eq:Isw}) simply becomes another way of writing $n_0^2 +
m_0^2 =1$.  In the thermodynamic limit, we transform Brillouin zone
sums to integrals according to
\begin{equation} 
  \frac{2}{N} \sum_{\bd k} \stackrel{N\to\infty}{\longrightarrow} 
  V_u \int_{\text{BZ}}\frac{d^2k}{(2\pi)^2}\,, 
\end{equation} 
where $V_u = a_1 a_2 \sin \varphi$ is the area of the magnetic unit
cell in real space and the integral is over the reduced Brillouin zone
shown in Fig.~\ref{fig:BZ}.

At $T=0$ and $h_s=0$ expressions similar to (\ref{eq:Fsw}) and
(\ref{eq:Isw}) have been discussed previously.\cite{Zhitomirsky98}
Only $m(h)$ was given explicitly and a renormalization of the canting
angle was found by considering spin-wave interactions. Yet, to a given
order in $1/S$ it is easier to calculate $m$ and $n$ directly as
derivatives of the free energy with respect to $h$ and $h_s$. Very
recently, the renormalized canting angle was also used to analyze the
behavior of $n(h)$ at $T=0$ for a more complicated
geometry.\cite{Veillette05}

At any finite temperature the integral $I(h,0)$ is infrared divergent
in two dimensions, signaling the absence of long-range
antiferromagnetic order, in accordance with the
Hohenberg-Mermin-Wagner theorem.\cite{Hohenberg67} At first sight,
it thus seems that the finite-temperature magnetization curve cannot
be calculated within our spin-wave approach.  Fortunately, there is a
straightforward way to obtain an approximate expression for the
magnetization even at finite $T$. The crucial observation is that if
we set $n=0$ in Eqs.~(\ref{eq:Fsw}) and (\ref{eq:Isw}), these equations can be
interpreted as a condition for the staggered field $h_s$ that is necessary to
enforce a vanishing staggered magnetization.  The solution
$h_s=h_s(h)$ as a function of the uniform field $h$ is not a physical
external staggered field, but an internal effective field that is
generated by strong fluctuations.  In fact, the field $h_s(h)$ is
nothing but the Lagrange multiplier introduced in Takahashi's modified
spin-wave theory.\cite{Takahashi89,Kollar03}  It is well known that
the internal field is related to a finite correlation length $\xi$, as
we will further discuss in Sec.~\ref{sec:stagg}.  Numerically, we
calculate the uniform magnetization $m(h,T)$ at finite temperature $T$
by adjusting $h_s$ for fixed external field $h$ such that the
condition $n=0$ is fulfilled in Eqs.~(\ref{eq:Fsw}) and
(\ref{eq:Isw}).  Using this $h_s(h)$ in Eq.~(\ref{eq:Fsw}) then
directly yields $m(h,T)$.

We must keep in mind that the staggered field $h_s$ does not respect
the rotational symmetry of the original hamiltonian, which for $h=0$
corresponds to a global O(3) symmetry and for $h>0$ is reduced to a
global O(2) symmetry around the axis of the uniform field. With the
parametrization that explicitly breaks this symmetry, we should
therefore only calculate rotationally invariant quantities.\cite{Kopietz97} 
Below, we
will find a disagreement between a rotationally invariant evaluation
of the zero-field uniform susceptibility and the slope of 
$\partial m/ \partial h$ for $h\to0$. 
We attribute this discrepancy to the fact that $\partial m/\partial h {\vert}_{h\to0}$
does not respect the O(3) symmetry in this limit. Generally, we expect
our approach for the finite temperature magnetization to be reasonable
only for $h>h_s(h,T)$. In Sec.~\ref{sec:stagg} we will see that $h_s$
is exponentially small at low temperatures, such that $h>h_s(h,T)$ is
fulfilled even for very small external fields. The condition
$h>h_s(h,T)$ then roughly gives a limit of validity of our approach in
terms of the temperature as $T\lesssim 0.5 \tilde{J}_0S$.
The fact that the limits $T \to 0$ and $h \to 0$ do not commute in a 
modified spin-wave expansion
was first noticed by Takahashi.\cite{Takahashi89}

\subsection{Uniform susceptibility}

In order to calculate the rotationally invariant uniform zero-field
susceptibility per spin
\begin{equation}
  \label{eq:chidef}
  \chi = \frac{1}{T N} \sum_{i,j} \langle \bd{S}_i \cdot \bd{S}_j \rangle\,,
\end{equation}
we set the uniform magnetic field $B=0$ in \mbox{Eq.(\ref{eq:hidef})}.
In this case $m=m_0=0$ and $n_0=1$, so that we obtain a doubly
degenerate mode in \mbox{Eq. (\ref{eq:epsilon2})} with dispersion
\begin{equation} 
  \epsilon_{\bd{k}\sigma}= \epsilon_{\bd{k}} = \sqrt{ (1+2h_s)^2 - |\gamma_{\bd{k}}|^2 }\,,
  \label{eq:epsilon0}
\end{equation}
and the expression for the staggered magnetization (\ref{eq:Isw})
reduces to
\begin{equation}
  \label{eq:I0SW}
  n = 1 + \frac{1}{2S} - 
  \frac{2}{NS}\sum_{\bd{k}}\frac{n_{\bd{k}}+\frac{1}{2}}{\epsilon_{\bd{k}}}
  \left( 1 + 2h_s \right)\,.
\end{equation}
As explained in the previous section we use a self-consistently
determined staggered field $h_s$ to enforce a vanishing order
parameter $n=0$.

The susceptibility (\ref{eq:chidef}) can be written as
\begin{equation}
  \label{eq:chi}
  \chi = \frac{1}{T}\left\langle\bd{S}_{\bd{q},+} \cdot \bd{S}_{-\bd{q},+}\right\rangle_{\bd{q}=0}\,,
\end{equation}
where we have defined the linear combinations ($\sigma=\pm$)
\begin{equation}
  \bd{S}_{\bd{q},\sigma} = \frac1{\sqrt{2}}\left(\bd{S}_{\bd{q}}^\mathrm{A}+
    \sigma\bd{S}_{\bd{q}}^\mathrm{B}\right)
\end{equation}
of the Fourier-transformed spin operators on each sublattice
\begin{equation}
  \label{eq:SFT}
  \bd{S}_{\bd{q}}^{\mathrm{A}/\mathrm{B}} = \sqrt{\frac{2}{N}}\sum_{\bd{r}_i\in \mathrm{A}/\mathrm{B}}e^{-i\bd{q}\cdot\bd{r}_i}\,\bd{S}_i\,.
\end{equation}
Next we decompose the susceptibility into a transverse and a
longitudinal part
\begin{equation}
  \label{eq:chitl}
  \chi = \chi^{+-} + \chi^{zz}\,,
\end{equation}
where 
\begin{eqnarray}
  \label{eq:chit}
  \chi^{+-}  & = & \frac{1}{2T} \left\langle 
  S^{+}_{\bd{q},+}S^{-}_{-\bd{q},+} + S^{+}_{\bd{q},+}S^{-}_{-\bd{q},+}
  \right\rangle_{\bd{q}=0}\,,\\
  \label{eq:chil}
  \chi^{zz}  & = & \frac{1}{T} 
  \left\langle S^{z}_{\bd{q},+}S^{z}_{-\bd{q},+}\right\rangle_{\bd{q}=0}\,.
\end{eqnarray}
We map the spin operators~(\ref{eq:SFT}) onto canonical boson 
operators via a Dyson-Maleev transformation\cite{Dyson56,Maleev57} 
and evaluate the thermal expectation 
values of the noninteracting state using the Wick theorem.
Then the  transverse susceptibility (\ref{eq:chit}) is
proportional to the right hand side of \mbox{Eq. (\ref{eq:I0SW})}, and
thus vanishes if we require $n=0$.
Therefore, in our approximation only the longitudinal
part contributes to the rotationally invariant uniform susceptibility,
\begin{equation}
  \label{eq:chisw}
  \chi =   \frac{2}{TN} \sum_{\bd{k}} n_{\bd{k}} (n_{\bd{k}}+1)\,.
\end{equation}
Apart from a different normalization,
this result has been obtained previously in Takahashi's
approach.\cite{Takahashi89}
We evaluate \mbox{Eq. (\ref{eq:chisw})} numerically in the thermodynamic limit.

\section{Results}
\label{sec:results}

\subsection{Zero temperature uniform and staggered magnetization}

\begin{figure}[tb]
  \begin{center}
    \psfrag{m}{$m$}
    \psfrag{h}{$h$}
    \psfrag{0}{\footnotesize{$0$}}
    \psfrag{0.2}{\footnotesize{$0.2$}}
    \psfrag{0.4}{\footnotesize{$0.4$}}
    \psfrag{0.6}{\footnotesize{$0.6$}}
    \psfrag{0.8}{\footnotesize{$0.8$}}
    \psfrag{1}{\footnotesize{$1$}}
    \psfrag{classical}{\footnotesize{classical}}
    \psfrag{square lattice}{\footnotesize{square lattice}}
    \psfrag{linear chain}{\footnotesize{linear chain}}
    \psfrag{honeycomb lattice, J1=J2}{\footnotesize{honeycomb lattice, $J_1/J_2=1$}}
    \psfrag{honeycomb lattice, J1=10J2}{\footnotesize{honeycomb lattice, $J_1/J_2=10$}}
    \epsfig{file=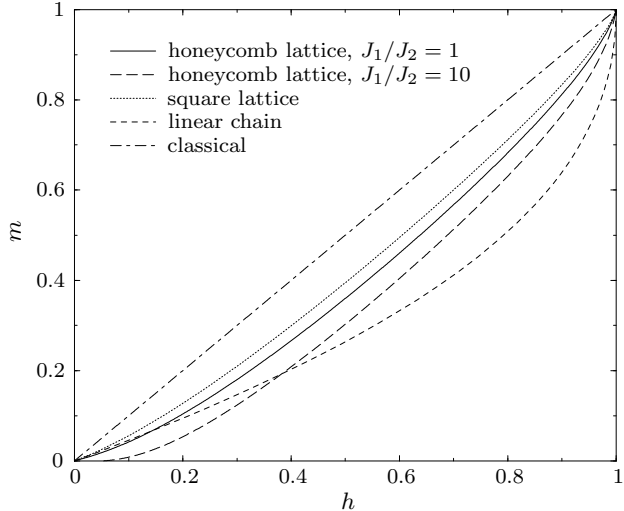,width=70mm,angle=270}~~~~~~

  \end{center}
  \vspace{-4mm}
  \caption{%
    Normalized uniform magnetization $m (h)$ for $T=0$ and $h_s=0$.
    The solid line is the zero-temperature magnetization curve for the
    honeycomb lattice with $S=1/2$ and $J_1 = J_2$.  For comparison we
    also show the corresponding curve for a square lattice and exact
    results for a linear antiferromagnetic chain.\cite{Griffiths64} 
    However, the $S=1/2$ chain is critical, so it is not surprising
    that it is poorly described by means of the spin-wave theory.
    Note that for $h_s=0$ the classical equation (\ref{eq:hh}) is simply
    $m_0=h$.}
  \label{fig:magn}
\end{figure}

\begin{figure}[tb]
  \begin{center}
    \psfrag{n}{$n$}
    \psfrag{h}{$h$}
    \psfrag{0}{\footnotesize{$0$}}
    \psfrag{0.2}{\footnotesize{$0.2$}}
    \psfrag{0.4}{\footnotesize{$0.4$}}
    \psfrag{0.6}{\footnotesize{$0.6$}}
    \psfrag{0.8}{\footnotesize{$0.8$}}
    \psfrag{1}{\footnotesize{$1$}}
    \psfrag{classical}{\footnotesize{classical}}
    \psfrag{square lattice}{\footnotesize{square lattice}}
    \psfrag{honeycomb lattice, J1=J2}{\footnotesize{honeycomb lattice, $J_1/J_2=1$}}
    \psfrag{honeycomb lattice, J1=10J2}{\footnotesize{honeycomb lattice, $J_1/J_2=10$}}
    \psfrag{honeycomb lattice, J1=100J2}{\footnotesize{honeycomb lattice, $J_1/J_2=100$}}
    \epsfig{file=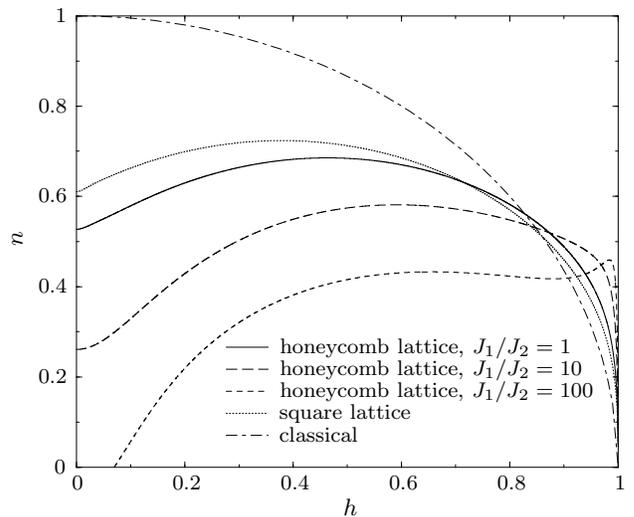,width=70mm,angle=270}~~~~~~
  \end{center}
  \vspace{-4mm}
  \caption{%
    Normalized staggered magnetization $n(h)$ at $T=0$ for honeycomb 
    (solid line) and square lattice with $S=1/2$ (dotted line). The 
    classical equation $n_0=\sqrt{1-h^2}$ is plotted for comparison. We
    also show the curves for the anisotropic cases $J_1/J_2=10,100$.}
  \label{fig:magnst}
\end{figure}

In Figs.~\ref{fig:magn} and \ref{fig:magnst} we show results for the
uniform and staggered magnetization at zero temperature. In two
spatial dimensions, there are no divergent contributions to the
integrals in Eqs.~(\ref{eq:Fsw}) and (\ref{eq:Isw}), indicating true
long range order.  We can thus set $h_s=0$ and consequently $m_0=h$.
As the deviations from the classical curves are rather small for
$S=5/2$, we present the curves for the extreme quantum case $S=1/2$.

The uniform magnetization shows a positive curvature for all $0\leq h
<1$ and lies generally below the classical straight
line.\cite{Zhitomirsky98} This tendency is stronger for the honeycomb
lattice and is even more pronounced for anisotropic exchange couplings
with $J_1\gg J_2$. The number of nearest neighbors $z=3$ for the
honeycomb lattice is lower than for the square lattice ($z=4$), and in
the limit $J_2\ll J_1$ the system is almost one-dimensional. The
observed tendency thus simply corresponds to increased quantum
fluctuations in low dimensions. Beyond the saturation field $h=1$ the
ground state has full collinear ferromagnetic order. This
state as well as single magnon excitations above it are easily shown
to be exact eigenstates. As the single magnon states become gapless at
exactly the classical value $h=1$, the saturation field is
not changed by quantum fluctuations or magnon interactions.  The limit
$h\to1$ is reached with infinite slope in $m(h)$. The leading behavior
is given by
\begin{equation}
  m =  1 + \frac{V_u}{4 \ell_x \ell_y}\frac{\delta h}{\pi S}
  \ln \left(4\delta h\right)\,,
\end{equation} 
where $\delta h=1-h$. This logarithmic asymptotics was first discussed in the
language of Bose condensation of magnons below the saturation
field\cite{Gluzman93} and was later found for the square lattice
($V_u/4\ell_x\ell_y$=1) within linear spin-wave
theory.\cite{Zhitomirsky98} For our distorted honeycomb lattice, we
have
\begin{equation}
  \frac{V_u}{\ell_x\ell_y}=\sqrt{\frac{(2J_1+J_2)^3}{J_1^2J_2}}\,,
\end{equation}
which diverges for $J_1\to0$ or $J_2\to0$ and thus exemplifies the
increasing deviations from the classical curve for strongly
anisotropic exchange couplings.

The staggered magnetization in Fig.~\ref{fig:magnst} shows a
non-monotonic dependence on the applied uniform field. For vanishing
$h$ the staggered magnetization decreases as we lower the effective
dimensionality. An external field apparently suppresses quantum
fluctuations and $n(h)$ first increases with $h$ before it reaches a
maximum and then vanishes for $h\to1$ with infinite slope. The
asymptotic behavior is given by
\begin{equation}
  n = - \frac{V_u}{2\ell_x \ell_y}\frac{\sqrt{\delta h}}{\pi S}
  \ln\left(4\delta h\right)\,.
\end{equation}
Interestingly, the quantum corrections to the staggered magnetization
are positive close to the saturation field and the spin-wave result
therefore intersects the classical curve. In a quasi one-dimensional
situation ($J_2\ll J_1$), quantum fluctuations are strong and the
leading order spin-wave theory, when pushed to the limit of validity,
predicts a quantum disordered phase for small uniform fields.

\subsection{Finite temperature magnetization and susceptibility}

Magnetic measurements were carried out on a single crystalline sample
of $\mathrm{Mn}[\mathrm{C}_{10}\mathrm{H}_{6}(\mathrm{OH})(\mathrm{COO})]_{2}
  \!\times\! 2\mathrm{H}_{2}\mathrm{O}$ with a
mass of $m_\mathrm{BONA}=0.65\,\mathrm{mg}$ using a Quantum Design SQUID
magnetometer MPMS-XL.  Isothermal magnetization runs at temperatures between
$2$ and $200\,\mathrm{K}$ and fields up to $5\,\mathrm{T}$ were
performed as well as measurements of the susceptibility in the
temperature range $2-300\,\mathrm{K}$ for a magnetic field of
$0.05-2\,\mathrm{T}$.\cite{Schmidt04}

\begin{figure}[tb]
  \begin{center}
    \psfrag{m}{$m$}
    \psfrag{h}{$h$}
    \psfrag{0}{\footnotesize{$0$}}
    \psfrag{0.2}{\footnotesize{$0.2$}}
    \psfrag{0.4}{\footnotesize{$0.4$}}
    \psfrag{0.6}{\footnotesize{$0.6$}}
    \psfrag{0.8}{\footnotesize{$0.8$}}
    \psfrag{1}{\footnotesize{$1$}}
    \psfrag{T=0}{$T=0$}
    \psfrag{T=0.5J0S}{$T=0.5\,\tilde{J}_0S$}
    \epsfig{file=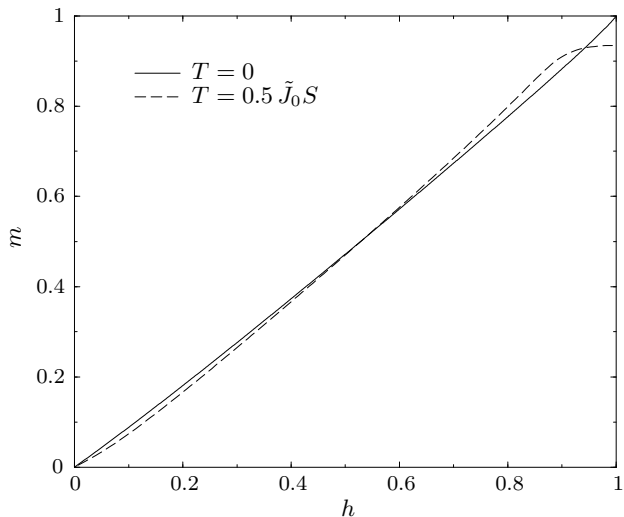,width=70mm,angle=270}~~~~~~
  \end{center}
  \vspace{-4mm}
  \caption{%
    Uniform magnetization $m (h)$ for the honeycomb lattice with
    $S=5/2$ and $J_1 = J_2$ for two values of $T$.  }
  \label{fig:magnT}
\end{figure}
In Fig.~\ref{fig:magnT} we show theoretical magnetization curves
$m(h)$ for the honeycomb lattice with $S=5/2$ and $J_1 = J_2$ at
different temperatures $T$.  For $T\ll\tilde{J}_0S$ the magnetization
is almost linear throughout the entire field range. At intermediate
temperatures $m(h)$ has an S-like shape with a positive curvature at
small fields $h$ that changes to a negative curvature with increasing
$h$.
Similar low-temperature behavior of the magnetization curve has been
observed in a quantum Monte Carlo study of the two-dimensional
Heisenberg antiferromagnet on a square lattice.\cite{Woodward02}

\begin{figure}[tb]
  \begin{center}
    \psfrag{m}{$m$}
    \psfrag{H[T]}{$H\,\mathrm{[T]}$}
    \psfrag{0}{\footnotesize{$0$}}
    \psfrag{0.1}{\footnotesize{$0.1$}}
    \psfrag{0.2}{\footnotesize{$0.2$}}
    \psfrag{0.3}{\footnotesize{$0.3$}}
    \psfrag{1}{\footnotesize{$1$}}
    \psfrag{2}{\footnotesize{$2$}}
    \psfrag{3}{\footnotesize{$3$}}
    \psfrag{4}{\footnotesize{$4$}}
    \psfrag{5}{\footnotesize{$5$}}
    \psfrag{1}{\footnotesize{$1$}}
    \psfrag{2K fit}{$2\,\mathrm{K}$ calculation}
    \psfrag{8K fit}{$8\,\mathrm{K}$ calculation}
    \psfrag{20K fit}{$20\,\mathrm{K}$ calculation}
    \psfrag{2K}{$2\,\mathrm{K}$ data}
    \psfrag{8K}{$8\,\mathrm{K}$ data}
    \psfrag{20K}{$20\,\mathrm{K}$ data}
    \epsfig{file=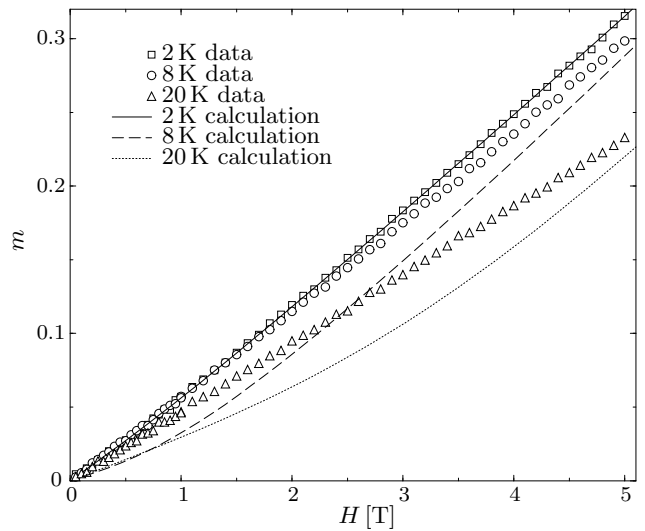,width=70mm,angle=270}~~~~~~
  \end{center}
  \vspace{-4mm}
  \caption{%
    Magnetization $m (H)$ of
    $\mathrm{Mn}[\mathrm{C}_{10}\mathrm{H}_{6}(\mathrm{OH})(\mathrm{COO})]_{2}
  \!\times\! 2\mathrm{H}_{2}\mathrm{O}$ up to
    field $H=5\,\mathrm{T}$.  Experimental data are indicated by
    squares ($2\,\mathrm{K}$), circles ($8\,\mathrm{K}$) and triangles
    ($20\,\mathrm{K}$).  Theoretical magnetization curves for
    honeycomb lattice with $S=5/2$ and $J_2=2 J_1=1.95\,\mathrm{K}$ are denoted by
    lines.  }
  \label{fig:magnfit}
\end{figure}

It turns out that the magnetization as well as the
susceptibility are not very sensitive to the ratio $J_2/J_1$ as long
as $J_1$ and $J_2$ have the same order of magnitude. Thus, we cannot
determine the precise value of $J_2/J_1$, but our fits are compatible
with the assumption $J_2 \approx 2J_1$.

In Fig.~\ref{fig:magnfit} we show experimental data and theoretical
fits for the normalized uniform magnetization $m=M/(NS)$ at different
temperatures.  The magnetic field $H=2\tilde{J}_0Sh$ is given in
Tesla. Surprisingly, all experimental curves are almost straight
lines, whereas from Fig.~\ref{fig:magnT} we would expect an upward
bend of $m(h)$ at higher temperatures. Fits for $T = 2\,
\mathrm{K}$ and different ratios $J_1/J_2$ invariably give
$\tilde{J}_0 \approx 4\, \mathrm{K}$.  Hence we assume $J_2 = 2 J_1$
and fit the theoretical curve to the experimental data at $T = 2\,
\mathrm{K}$.  Good agreement is achieved for $J_2 = 1.95\,
\mathrm{K}$. For this value of the exchange couplings, we also plot
theoretical magnetization curves at $T = 8\,\mathrm{K}$ and $T=20\,
\mathrm{K}$ in Fig.~\ref{fig:magnfit}. These curves deviate
significantly from the data, but one should be aware that $T=8K$ is
already beyond the estimated limit of validity $T\lesssim 0.5
\tilde{J}_0S$ of our theoretical approach. 

\begin{figure}[tb]
  \begin{center}
    \psfrag{T[K]}{$T\,\mathrm{[K]}$}
    \psfrag{chi}{\!\!\!\!\!\!\!$\chi\,\mathrm{[cm^3/mol]}$}
    \psfrag{0}{\footnotesize{$0$}}
    \psfrag{0.05}{\footnotesize{$0.05$}}
    \psfrag{0.1}{\footnotesize{$0.1$}}
    \psfrag{0.15}{\footnotesize{$0.15$}}
    \psfrag{100}{\footnotesize{$100$}}
    \psfrag{200}{\footnotesize{$200$}}
    \psfrag{300}{\footnotesize{$300$}}
    \epsfig{file=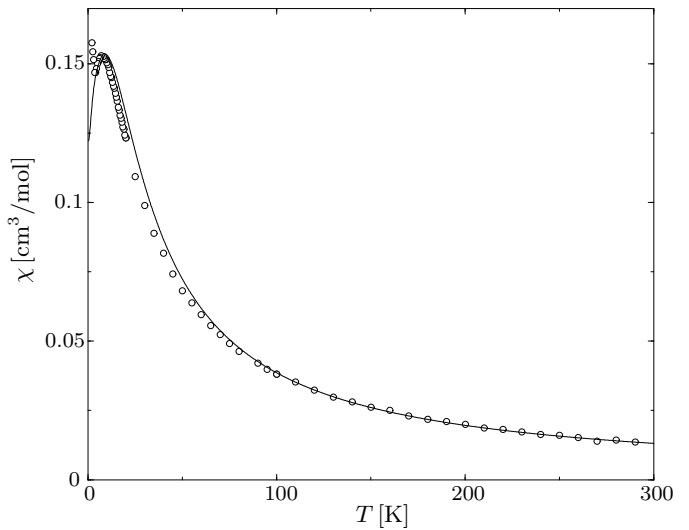,width=70mm,angle=270}~~~~~~
  \end{center}
  \vspace{-4mm}
  \caption{%
    Susceptibility $\chi (T)$ of
    $\mathrm{Mn}[\mathrm{C}_{10}\mathrm{H}_{6}(\mathrm{OH})(\mathrm{COO})]_{2}
  \!\times\! 2\mathrm{H}_{2}\mathrm{O}$.
    Circles are experimental data in a field of $2\,\mathrm{T}$.
    Theoretical fit for honeycomb lattice with $J_2=2 J_1$ (solid
    line) gives $J_2=1.66K$.}
  \label{fig:suscfit}
\end{figure}

In Fig.~\ref{fig:suscfit} the uniform susceptibility is plotted in the
experimental units $\mathrm{cm^3/mol}$. When all exchange integrals
have the same order of magnitude we expect a peak in the
susceptibility for $T \approx \tilde{J}_0 S$.  Experimentally, the
peak is at approximately $7\, \mathrm{K}$ so that we have
$\tilde{J}_0\approx 3\, \mathrm{K}$, in accordance with the fits of
the magnetization curves.  For a more quantitative comparison we use the
following procedure. First we substract the temperature-independent
contribution from the
experimental susceptibility in order to get the correct paramagnetic
behavior at high temperatures.  Then we fit the theoretical expression
(\ref{eq:chisw}) with $J_2 = 2 J_1$ to the full set of data points.
Circles in Fig.~\ref{fig:suscfit} are experimental data and the solid
line is a fit with $J_2 = 1.66\, \mathrm{K}$.  The theoretical curve
reproduces the behavior of the susceptibility very well and it
especially gives a good estimate of the position and the form of the
peak. Note that we experimentally observe an increase in the
susceptibility below $T_{\ast} = 3.0 \pm 0.2 \mathrm{K}$. This coincides with an
anomaly in the specific heat.
The careful reader will notice at this point that the estimated value 
of $T_{\ast}$ is larger than the temperature $T = 2\, \mathrm{K}$ where we obtained
the best fit of our calculated magnetization curve $m (H)$ to the experimental
data shown in Fig.~\ref{fig:magnfit}. Hence, at $T = 2\, \mathrm{K}$ the system 
seems to have some kind of long range magnetic order, which we have ignored in our 
calculation. However, the precise nature of the order and the mechanism 
responsible for the ordering are not known at this point. The fact that a
strictly 2D model can reasonably well explain the 
magnetization curve at $T = 2\, \mathrm{K}$
imposes some constraint on possible ordering mechanisms. We suspect that
dipole-dipole interactions play an important role in this temperature range, 
because the long-range nature of the dipole-dipole interaction can give
rise to spontaneous antiferromagnetic order even in 2D.\cite{Pich93} 
This point deserves further attention, both theoretically and experimentally.


\subsection{Staggered correlation length in a magnetic field}
\label{sec:stagg}

The energy gap appearing in Eq.~(\ref{eq:epsilonminuslong}) can be
related to the staggered correlation length $\xi$, as discussed by
Takahashi.\cite{Takahashi89} Assuming for simplicity $ |
\bd{\delta}_{\nu} | = a$, we may identify
\begin{equation} 
  \left(\frac{a}{2 \xi}\right)^2 = \Delta^2 = \frac{4h_s}{n_0}\,.
  \label{eq:xidef} 
\end{equation} 
In the absence of a uniform field the low temperature behavior of
$\xi$ has been thoroughly studied by Chakravarty, Halperin and
Nelson.\cite{Chakravarty88} Surprisingly, the effect of a uniform
field $h$ on $\xi$ has so far not been investigated.  We now analyze
the asymptotic behavior of $\xi$ at low temperatures.  In two spatial
dimensions, the limit $T\to0$ also implies $h_s\to0$.  Our
self-consistency equations (\ref{eq:Fsw}) and (\ref{eq:Isw}) can then
be solved analytically by isolating divergent contributions to the
integrals $I ( h, h_s)$ and $F( h, h_s)$ originating from gapless 
modes in the spin-wave spectrum.  In the regular part of the integral,
the limit $T\to0$ and $h_s\to0$ can be taken. For the leading behavior
at small uniform fields $h\ll 1$ only the singular part of $I(h,h_s)$ 
contributes, and we obtain the self-consistency condition
\begin{equation}
  0 = n(0) - \frac{I^{\text{sing}}(h,h_s)}{S}\,.
  \label{eq:selfcon}
\end{equation}
Here, $I^{\text{sing}}(h,h_s)$ is the part of the integral $I (
h,h_s)$ that diverges for vanishing gaps in the spin-wave dispersions,
and $n(0) = n (h=0, h_s=0, T=0)$.  For $h\ll1$, we obtain
\begin{eqnarray}
  I^{\text{sing}}(h,h_s)&  = & \frac{T}{\tilde{J}_0 S} \frac{V_u}2 \sum_{\sigma}
  \int \frac{d^2k}{(2\pi)^2}
  \frac{1}{\epsilon_{{\bd k}\sigma}^2}
  \nonumber
  \\  
  & \approx & - \frac{T}{ \tilde{J}_0 S }\frac{V_u}{8 \pi \ell_x \ell_y}
  \Bigg[\ln\left(\frac{4h_s}{n_0}\right)
  \nonumber\\[1mm]
  &&+ \ln\left(4h^2+\frac{4h_s}{n_0}\right)
  \Bigg]\,,
  \label{eq:Idiv}
\end{eqnarray}
to leading logarithmic order. From Eqs.~(\ref{eq:xidef}) and
(\ref{eq:selfcon}) we then obtain the following result for the
self-consistent energy gap in a small uniform magnetic field
\begin{equation} 
  \Delta^2(h) = \left(\frac{a}{2\xi(h)}\right)^2
  =\sqrt{\Delta_0^4+\frac{(2h)^4}4}-\frac{(2h)^2}2\,,
  \label{eq:Deltah}
\end{equation}
where $\Delta_0=a/2\xi(0)$ is the gap for vanishing uniform field and
the temperature dependence of the zero-field staggered correlation
length is given by
\begin{equation} 
  \frac{\xi(0)}{a} \propto \exp\left(
    \frac{ 2\pi \tilde{J}_0  S^2 n (0) }{T}
    \frac{\ell_x \ell_y}{V_u}
  \right)\,.
  \label{eq:xi0}
\end{equation}
For a square lattice this yields with $\tilde{J}_0 = 4 J$ and $\ell_x
\ell_y / V_u = 1/4$
\begin{equation}
  \frac{\xi(0)}{a} \propto \exp\left(
    \frac{2  \pi J  S^2 n (0) }{T} 
  \right)
  \,,
  \label{eq:xi0square}
\end{equation}
which is identical to Takahashi's result (see Eq.~(27a) in
Ref.~\onlinecite{Takahashi89}), except that we do not include a
spin-wave velocity renormalization in our approach.  To obtain this
renormalization, the spin-wave interaction would have to be treated on
the mean-field level in a fully self-consistent way.

The field dependence of the correlation length for fixed temperature is
given by Eq.~(\ref{eq:Deltah}). For $h\ll\Delta_0(T)$, we have
\begin{equation}
  \xi(h)=\xi(0)\left[1+\frac12\left(\frac{h}{\Delta_0}\right)^2\right]\,,
\end{equation}
whereas for $h\gg\Delta_0(T)$, we obtain
\begin{equation}
  \frac{\xi(h)}a = 4h\,\left(\frac{\xi(0)}a\right)^2\,.
\end{equation}
From Eq.~(\ref{eq:Deltah}) it is clear that $\xi(h)>\xi(0)$. Thus, the
correlation length is increased by a small uniform field due to reduced 
quantum fluctuations. 

The temperature dependence of the correlation length for fixed uniform
field $h$ can also be extracted from Eq.~(\ref{eq:Deltah}).  As long
as $\Delta_0(T) \gg 2h$, this temperature dependence is still given by
Eq.~(\ref{eq:xi0}). When the temperature is further reduced, 
Eq.~(\ref{eq:Deltah}) predicts a crossover at $\Delta_0(T) \approx 2h$
to the following temperature-dependent correlation length
\begin{equation}
  \frac{\xi (h)}{a} \propto \exp\left(
    \frac{4  \pi \tilde{J}_0  S^2 n(0)}{T}
    \frac{\ell_x \ell_y}{V_u}
  \right)\,.
  \label{eq:xih}
\end{equation}
The additional factor of two in the exponent as compared to
Eq.~(\ref{eq:xi0}) is due to the fact that at very low temperatures
the spin-wave mode $\epsilon_{{\bm k}-}$ yields a singular
contribution, whereas the mode $\epsilon_{{\bm k}+}$ has a gap $~2h$
which is fixed by the external field. In contrast, for $h=0$ both
modes contribute equally, leading to Eq.~(\ref{eq:xi0}). 

The analysis in this section has been carried out for $h\ll1$. For
larger fields, there are field dependent prefactors of the first
logarithm in Eq.~(\ref{eq:Idiv}) leading to a field dependent
renormalization factor $Z_h$ in the exponent of Eq.~(\ref{eq:xih}).
The field dependence of the correlation length at fixed temperature is
then no longer determined by the singular contributions to the
integrals and cannot be extracted from the simple analysis presented
here. Close to the critical field at $h=1$ the nature of the
divergences changes, since the dispersion of the $\sigma=-$ mode
becomes quadratic. As our mean-field calculation is not suitable to
describe the true critical behavior in two dimensions, we do not
discuss this limit in more detail.

Our approach can also describe a quasi one-dimensional anisotropic
system, where the exchange coupling between chains is very weak. The
dispersion is then almost flat in the transverse direction. The
integrals will be quasi one-dimensional as long as the maximum
variation of the dispersion in the transverse direction is smaller
than the self-consistent gap $4h_s /n_0$. In this intermediate
temperature regime the staggered correlation length behaves as if the
system were one-dimensional.  At even lower temperatures there will be
a crossover to the true asymptotic two-dimensional behavior. A rough
estimate for the position of the crossover is ${a}/{\xi} \propto
{\ell_{\bot}}/{a}$ where $\ell^2_{\bot}$ is the eigenvalue of the
matrix $\bd{A}$ defined in Eq.~(\ref{eq:Amatrixdef}) associated with
the eigenvector perpendicular to the chain direction.
 
\section{Conclusion}
\label{sec:discussion}

In summary, we have investigated the magnetic properties of the new
metal-organic quantum magnet
$\mathrm{Mn}[\mathrm{C}_{10}\mathrm{H}_{6}(\mathrm{OH})(\mathrm{COO})]_{2}
  \!\times\! 2\mathrm{H}_{2}\mathrm{O}$.
Its layered structure contains two-dimensional arrangements of
$\mathrm{Mn}^{2+}$ ions that suggest a spin $S=5/2$ Heisenberg model on a
distorted honeycomb lattice as a minimal model. In order to explain
measurements of the magnetization $M(H,T)$ and the
susceptibility $\chi(T)$, we develop a variant of modified spin-wave
theory, which can be used to describe finite temperature properties of
two-dimensional magnets in a uniform external magnetic field.  A fit
of the theoretical results to the experimental curves shows a
satisfactory agreement for the magnetization at low temperatures
where we expect our theoretical approach to be valid. The magnetic
susceptibility is very well described down to temperatures of
$T \gtrsim T_{\ast} \approx 3\mathrm{K}$. 
Both quantities are consistently fitted by one parameter 
$J_2 = 2 J_1$ to give the exchange coupling 
$J_2 \approx 1.8\,
\mathrm{K}$. For temperatures below $T_{\ast}$ the uniform susceptibility
shows again an upturn, which together with an anomaly in the specific
heat is most likely due to some ordering transition. Possible
mechanisms for this transition are dipole-dipole interactions or
couplings between the layers, which should be included in more refined
theoretical models. From the experimental point of view  
nuclear magnetic resonance or neutron scattering  measurements 
could provide a more detailed insight into the nature of the 
magnetic interactions.

This work was supported by the DFG via Forschergruppe FOR 412.
We thank M.~Kuli\'{c} for interesting discussions and especially for 
pointing out the possible importance of
dipole-dipole interactions.


\end{document}